\documentstyle[12pt]{article}
\newlength{\mytopmargin}
\newlength{\myleftmargin}
\def\zz{\rlx\hbox{\small \sf Z\kern-.4em Z}}

\begin{document}

\vspace{1cm}
\noindent
\begin{center}{\large \bf  Fluctuation formula for complex random matrices}
\end{center}
\vspace{5mm}

\noindent
{\center 
 P.J.~Forrester
\\
\it Department of Mathematics, University of Melbourne, Parkville, Victoria
3052, Australia}
\vspace{.5cm}

\small
\begin{quote} A Gaussian fluctuation formula is proved for linear
statistics of complex random matrices in the case that the statistic
is rotationally invariant. For a  general linear statistic without
this symmetry, Coulomb gas theory is used to predict that the
distribution will again be a Gaussian, with a specific mean and
variance. The variance splits naturally into a bulk and surface
contibution, the latter resulting from the long range correlations
at the boundary of the support of the eigenvalue density.
\end{quote}

\vspace{.5cm}
The phenomenum of universal conductance fluctuations in mesoscopic
wires (see e.g.~\cite{Be97}) has provided the motivation for a number
of theoretical studies into fluctuation formulas for linear statistics
in random matrix ensembles \cite{2}. To
understand the reason for this, we first recall that the striking feature
of the conductance fluctuations is that they remain of order unity even
though the conductance itself is proportional to the number of channels
$N$. Now, in random matrix models of this effect, the conductance can
be written as a linear statistic of a certain random matrix ensemble
(we recall that $A$ is said to be linear statistic of the
eigenvalues $\lambda_j$ if it can be written in the form
$A = \sum_{j=1}^n a(\lambda_j)$ for some random function $a$). In
this setting, the theoretical explanation for the phenomenum of
universal conductance fluctuations is as an example of a universal
fluctuation formula in random matrix theory, the first example of which
was given in the pioneering work of Dyson and Mehta \cite{DM63}.

For random matrix ensembles in which the support of the density is
one-dimensional, for example Hermitian or unitary random matrices,
(Gaussian) fluctuation formulas are now well understood both at a
heuristic (see references cited above) and rigorous level 
\cite{4,BF97f,6}. It is the purpose of this
Letter to initiate the study of fluctuation formulas in
complex random matrices \cite{Gi65,FKS97,9}, for which the eigenvalues
uniformly fill a disk or ellipse in the complex plane. We remark that
complex random matrices have occured in recent physical studies of
the localization-delocalization transition in non-Hermitian quantum
mechanics \cite{Ef97} and chiral symmetry breaking in lattice QCD
\cite{St96}. 
The distribution of a linear statistic is then an observable
quantity after averaging over many random copies of these
systems.

To begin, we recall \cite{Gi65} that for a random matrix with complex
elements $u_{jk} + i v_{jk}$ independently distributed with Gaussian
distribution ${c \over \pi} e^{-c(|u_{jk}|^2 + |v_{jk}|^2)}$, the
corresponding probability distribution of the eigenvalues
$\lambda_j = x_j + i y_j$ is proportional to
\begin{equation}\label{1}
\prod_{j=1}^N e^{-c|\vec{r}_j|^2} 
\prod_{1 \le j < k \le N} |\vec{r}_j - \vec{r}_k |^2,
\end{equation}
where $\vec{r}_j = (x_j,y_j)$. Furthermore, to leading order, the support
of the density of the eigenvalues is the disk of radius $\sqrt{N/c}$.
For the purpose of studying fluctuation formulas it is convenient
to choose $c=N$ so that the support of the density is the unit disk.
The  Fourier transform of the distribution of
Pr$(A=u)$ is then given by
\begin{equation}\label{2}
\tilde{P}(k) =
{\prod_{l=1}^N \int_{{\bf R}^2} d \vec{r}_l \,
e^{-N |\vec{r}_l|^2 + i k a(\vec{r}_l)} \prod_{1 \le j < k \le N}
|\vec{r}_k - \vec{r}_j|^2 \over
\prod_{l=1}^N \int_{{\bf R}^2} d \vec{r}_l \,
e^{-N |\vec{r}_l|^2 } \prod_{1 \le j < k \le N}
|\vec{r}_k - \vec{r}_j|^2}.
\end{equation}

Suppose now that the linear statistic is of the form 
$A = \sum_{j=1}^N a(|\vec{r_j}|)$, so that the random function
$a$ only depends on the distance from the origin. Introducing polar
coordinates and using the Vandermonde determinant expansion of
$\prod_{1 \le j < k \le N}(z_k - z_j)$ the integrals in 
(\ref{2}) can be
evaluated with the result
\begin{equation}\label{3}
\tilde{P}(k) = {\prod_{l=1}^N \int_0^\infty e^{-s} s^{l-1}
e^{i k a(\sqrt{s/N})} \, ds \over
\prod_{l=1}^N \int_0^\infty e^{-s} s^{l-1} \, ds}.
\end{equation}
This is the exact expression for finite $N$. To obtain its form for
$N \to \infty$, we change variables $s \to ls$ and expand the
integrand about its large-$l$ maximum at $s=1$. A straightforward
calculation then gives
\begin{equation}
\tilde{P}(k)  \sim  \prod_{l=1}^N e^{ik a(\sqrt{l/N})}
e^{-k^2(a'(\sqrt{l/N}))^2}
 \sim  e^{ik\mu} e^{-k^2 \sigma^2/2}
\end{equation}
with
\begin{equation}\label{4}
\mu = 2N \int_0^1 r a(r) \, dr, \quad \sigma^2 = {1 \over 2} \int_0^1
r(a'(r))^2 \, dr.
\end{equation}
Thus the distribution of $A$ is a Gaussian with mean and variance as
given by (\ref{4}). Note in particular that the variance is $O(1)$.

Next we address the more general situation in which $a$ is not
rotationally invariant. To make progress we must proceed
heuristically. The p.d.f.~(\ref{1}) can be interpreted as the Boltzmann
factor of the two-dimensional one-component plasma (2dOCP) at the
special value of the coupling $\Gamma = 2$ \cite{AJ81}.
Using linear response theory and macroscopic electrostatics, it is
possible to argue \cite{Po89,Ja95} that in general the distribution of a
linear statistic in a classical Coulomb system in the conductive phase
will be Gaussian (this assumes also that the random function
varies over macroscopic distances relative to the interparticle spacing).
The Gaussian distribution is uniquely characterized by its mean and
variance. But independent of the underlying distribution, these
quantities are given by
\begin{equation}\label{5}
\mu = {N \over \pi} \int_{\Lambda} d\vec{r} \, a(\vec{r}), \quad
\sigma^2 =  \int_{\Lambda} d\vec{r}_1 a(\vec{r}_1) 
\int_{\Lambda} d\vec{r}_2 a(\vec{r}_2) S(\vec{r}_1,\vec{r}_2),
\end{equation}
where $S(\vec{r}_1,\vec{r}_2)$ is given in terms of the truncated
two particle distribution function by $S(\vec{r}_1,\vec{r}_2) =
\rho_{(2)}^T(\vec{r}_1,\vec{r}_2) + {N \over \pi} \delta(\vec{r} -\vec{r}\,')$,
${N \over \pi}$ is the particle density and $\Lambda$ denotes the unit
disk. We see immediately from the formula for $\mu$ in (\ref{5})
that the formula for $\mu$ in (\ref{4}) 
is reclaimed if $a(\vec{r}) = a(|\vec{r}|)$. More challenging is to
reproduce the formula for $\sigma^2$, and to proceed to generalize this
formula for general $a(\vec{r})$.

This task can be undertaken by again appealing to Coulomb gas theory.
In the infinite density limit the function $S(\vec{r}_1,\vec{r}_2)$
in (\ref{5}) for the 2dOCP with general coupling 
$\Gamma$ is expected to have the {\it bulk}
universal
form \cite{Ja95}
\begin{eqnarray}\label{ee}
S_{\rm bulk}(\vec{r}_1,\vec{r}_2) & = & - {1 \over 2 \pi \Gamma}
\nabla^2 \delta(\vec{r}_1 - \vec{r}_2) \nonumber \\ 
& = & {1 \over 2 \pi \Gamma}
\Big ( {\partial \over \partial x^{(1)}} + i
{\partial \over \partial y^{(1)}} \Big )
\Big ( {\partial \over \partial x^{(2)}} - i
{\partial \over \partial y^{(2)}} \Big ) \delta(\vec{r}_1 - \vec{r}_2).
\end{eqnarray}
At $\Gamma = 2$ this can be checked from the $\rho \to \infty$ limit
of the exact formula \cite{Gi65}
$S_{\rm bulk}(\vec{r}_1,\vec{r}_2) =
- \rho^2 e^{-\pi \rho |\vec{r}_1-\vec{r}_2|^2} + \rho 
\delta(\vec{r}_1-\vec{r}_2)$.
Substituting (\ref{ee}) in (\ref{5}), and integrating by parts 
(ignoring possible boundary terms, which are separately treated below) gives
\begin{equation}\label{last}
\sigma_{\rm bulk}^2 = {1 \over 2 \pi \Gamma} \int_{\Lambda} dx dy \,
\bigg ( \Big ( {\partial a(x,y) \over \partial x} \Big )^2 + 
 \Big ( {\partial a(x,y) \over \partial y} \Big )^2 \bigg ).
\end{equation}
In the special case $a(\vec{r}) = a(|\vec{r}|)$, $\Gamma = 2$, (\ref{last})
reproduces the result (\ref{4}) for $\sigma^2$.

The crucial difference between the case of general $a(\vec{r})$ and the
rotationally invariant case $a(\vec{r}) = a(|\vec{r}|)$ is that in the 
latter case $\sigma^2$ contains a contribution from the surface
correlations of the same order ($O(1)$) as the contribution from the
bulk correlations. This effect, due to the long-range nature of the
correlations at the boundary of Coulomb systems \cite{Ja82}, was
first noted by Choquard \cite{CPR86} and collaborators, who
studied the variance of the dipole moment ($a(\vec{r}) = x$) for
classical Coulomb systems. Indeed the variance of this statistic was
used to compute from microscopic statistical mechanics the macroscopic
{\it shape dependent} dielectric susceptibility of the Coulomb system.

Like in the bulk, the correlation $S(\vec{r}_1,\vec{r}_2)$ has a
universal form for $\vec{r}_1$ and $\vec{r}_2$ at the surface. However,
unlike the situation in the bulk, this correlation is long-ranged and
shape dependent. For $\Lambda$ a unit disk, the universal form is
\cite{Ja95}
\begin{equation}\label{1c}
S_{\rm sur.}((r_1,\theta_1),(r_2,\theta_2)) =
{2 \over \pi^2 \Gamma} \bigg (
{\partial^2 \over \partial \theta_1 \partial \theta_2}
\log \Big | \sin {\theta_1 - \theta_2 \over 2} \Big | \bigg )
\delta(r_1 - 1) \delta(r_2 - 1),
\end{equation}
where polar coordinates have been introduced. 
At $\Gamma = 2$ this form can be derived explicitly from the exact
evaluation of $S(\vec{r}_1,\vec{r}_2)$ in the finite system
\cite{CPR86}.
Substituting in (\ref{5})
gives
\begin{eqnarray}\label{1d}
\sigma_{\rm surface}^2 & = &
{2 \over \pi^2 \Gamma} \int_0^{2 \pi} d\theta_1
\Big ( {\partial \over \partial \theta_1}a(1,\theta_1) \Big )
 \int_0^{2 \pi} d\theta_2
\Big ( {\partial \over \partial \theta_2}a(1,\theta_2) \Big )
\log \Big | \sin {\theta_1 - \theta_2 \over 2} \Big | \nonumber \\
& = & {2 \over \Gamma} \sum_{n=1}^\infty n |a_n|^2, \qquad
a(1,\theta) = \sum_{n=-\infty}^\infty a_n e^{i n \theta}.
\end{eqnarray}
This quantity vanishes for $a(\vec{r}) = a(|\vec{r}|)$.

Consider now a more general ensemble of complex random matrices
\cite{FKS97}, in which the members, $J$ say, are of the form $J=H + ivA$.
Here $H$ and $A$ are Gaussian Hermitian random matrices with joint
p.d.f.'s for the elements proportional to $\exp\Big ( - {N \over
1 + \tau} {\rm Tr}X^2 \Big )$ ($X = H,A$ and $\tau = (1-v^2)/(1+v^2)$).
The corresponding eigenvalue p.d.f.~is proportional to
\begin{equation}\label{1e}
\exp \Big ( - N \sum_{j=1}^N ({x_j^2 \over 1 + \tau} +
{y_j^2 \over 1 - \tau}) \Big ) \prod_{1 \le j < k \le N}
|\vec{r}_j - \vec{r}_k|^2
\end{equation}
(note that in the case $\tau = 0$ (\ref{1e}) agrees with (\ref{1})), and to
leading order the support of the eigenvalue density consists of an
ellipse with semi-axes $A=(1+\tau)$, $B=(1-\tau)$. The eigenvalue density
itself is uniform and thus has the value $N/(\pi(1-\tau^2)$ inside the
ellipse.

The p.d.f.~(\ref{1e}) can be interpreted as the Boltzmann factor of
the 2dOCP at $\Gamma = 2$ in a quadrupolar field \cite{DGIL94,FJ96}. Thus,
Coulomb gas theory gives that as $N \to \infty$ (infinite
density limit) the distribution of
a linear statistic will again be Gaussian. The mean will be given as in
(\ref{5}), but with $\Lambda$ now the ellipse specifying the support of
the eigenvalues, and the factor $N/\pi$ which represents the eigenvalue
density replaced by $N/(\pi(1 - \tau^2))$. With $\Lambda$  the ellipse,
the bulk contribution to the variance is again given by (\ref{last}).
For the surface contribution, we require the fact that the
universal form of the surface correlation at the boundary of an
ellipse is \cite{CPR86,FJ96}
\begin{equation}\label{1f}
S_{\rm sur.}((\xi_1,\eta_1),(\xi_2,\eta_2) =
{2 \over \pi^2 \Gamma h(\xi_b,\eta_1)  h(\xi_b,\eta_2)}
\bigg ( {\partial^2 \over \partial \eta_1 \partial \eta_2}
\log \Big | \sin {\eta_1 - \eta_2 \over 2} \Big | \bigg )
\delta(\xi_1 - \xi_b) \delta(\xi_2 - \xi_b),
\end{equation}
where $(\xi,\eta)$ are elliptic coordinates, specified by $x+iy =
\cosh(\xi + i \eta)$, and $h(\xi_b,\eta) d \eta$ gives the
differential surface element. 
As in the disk case (\ref{1c}), this form has been explicitly verified
\cite{FJ96} from the known exact expression for $S(\vec{r}_1,
\vec{r}_2)$ in the finite system.
Since the semi-axes are specified by
$A = \cosh \xi_b$, $B = \sinh \xi_b$, $\xi_b$ is related to $\tau$ by
$\tanh \xi_b = (1 - \tau)/(1 + \tau)$. Substituting (\ref{1f}) in
(\ref{5}) gives
\begin{eqnarray}\label{1g}
\sigma_{\rm surface}^2 & = &
{2 \over \pi^2 \Gamma} \int_0^{2 \pi} d\eta_1
\Big ( {\partial \over \partial \eta_1}a(\xi_b,\eta_1) \Big )
 \int_0^{2 \pi} d\eta_2
\Big ( {\partial \over \partial \eta_2}a(\xi_b,\eta_2) \Big )
\log \Big | \sin {\eta_1 - \eta_2 \over 2} \Big | \nonumber \\
& = & {2 \over \Gamma} \sum_{n=1}^\infty n |a_n|^2, \qquad
a(\xi_b,\eta) = \sum_{n=-\infty}^\infty a_n e^{i n \eta}.
\end{eqnarray}
(note the similarity between (\ref{1g}) and (\ref{1d})).

There is a simple linear statistic for which the exact distribution
can be calculated, thus allowing the above predictions to be tested.
This statistic is the linear function
$a(x,y) =  c_{10}x + c_{01} y$.
Substituting  in the analogue of (\ref{2}) for the p.d.f.~(\ref{1e})
raised to the power $\Gamma / 2$,
the resulting dependence on $k$
can be simply calculated by completing the square and changing
variables in the integrand (see \cite{BF97f} for an analogous
calculation in the case of Hermitian random matrices). We find
\begin{equation}\label{t}
\tilde{P}(k) = e^{-k^2(c_{10}^2(1+\tau) + c_{01}^2(1 - \tau))/(2
\Gamma)},
\end{equation}
independent of $N$. Thus $\sigma^2 = {1 \over \Gamma}(c_{10}^2(1 + \tau) +
c_{01}^2(1 - \tau))$. Substituting the linear function
$a(x,y)$ in
(\ref{last}) and (\ref{1g}) and adding the result verifies that the
general formulas reproduce the exact value.

\noindent
{\bf Acknowledgement} \quad I thank A.~Soshnikov for detailed discussions,
and M.~Kiessling for comments on the manuscript. 
This work was supported by the Australian Research Council.

\end{document}